\documentclass{article}
\usepackage{spconf,amsmath,epsfig}
\graphicspath{{/media/bagustris/atmaja/s3/mypaper/2020/SPro2020/LaTeX/}}
\usepackage{url}

\let\OLDthebibliography\thebibliography
\renewcommand\thebibliography[1]{
  \OLDthebibliography{#1}
  \setlength{\parskip}{0pt}
  \setlength{\itemsep}{0pt plus 0.3ex}
}

\begin{document}\sloppy

\def\x{{\mathbf x}}
\def\L{{\cal L}}

\title{The Effect of Silence Feature in Dimensional Speech Emotion Recognition}
%
\name{Bagus Tris Atmaja$^{1,2}$, Masato Akagi$^{2}$}
\address{$^1$ Sepuluh Nopember Instiute of Technology, Surabaya, Indonesia \\
         $^2$ Japan Advanced Instiute of Science and Technology, Nomi, Japan\\
         Email: \url{bagus@ep.its.ac.id}}

\maketitle

\begin{abstract}
Silence is a part of human-to-human communication, which can be a clue for human emotion perception. For automatic emotion recognition by a computer, it is not clear whether silence is useful to determine human emotion within a speech. 
This paper presents an investigation of the effect of using silence feature in dimensional emotion recognition. Since the silence feature is extracted per utterance, we grouped the silence feature with high statistical functions from a set of acoustic features. The result reveals that the silence features affect the arousal dimension more than other emotion dimensions. The proper choice of a threshold factor in the calculation of silence feature improved the performance of dimensional speech emotion recognition performance, in terms of a concordance correlation coefficient. On the other side, improper choice of that factor leads to a decrease in performance by using the same architecture.
\end{abstract}
\noindent\textbf{Index Terms}: speech emotion recognition, dimensional emotion, silence feature, silence threshold, affective computing

\section{Introduction}
\label{sec:intro}

One of the elements of human to computer communication is the perception, which is implemented as automatic recognition in computers. Perception is the application's ability to consume, organize, and classify information about the user's physical and digital, and current and historical context. Perceptual data includes things like location, date, time, mood, expression, environment, physiological responses, connected applications, networks, and nearby devices \cite{ibm_developer_2018}. Due to this difference with human communication, especially on processing the data, the processing mechanism to obtain perceptual data on human--to--machine communication may be different from human--to--human communication.

Emotion is one of human perceptions. The difference between emotion and mood is that emotions are short-lived feelings that come from a known cause, while moods are feelings that are longer lasting than emotions and often 
without apparent cause \cite{scherer2000psychological}. Emotions can range from happy, ecstatic, sad, and prideful in the category, while moods are either positive or negative. Emotion also can be described in a degree of valence, arousal, and dominance. Other researchers used liking \cite{ringeval2019avec} and expectancy \cite{moore2014word} as additional dimensions or attributes to those dimensional emotions.

Valence (V) is the pleasantness of the stimulus [pleasure (P)], ranges from positive (extreme happy) to negative (extreme unhappy). In other words, it is also known as ``sentiment" or ``semantic orientation" \cite{jurafsky2018speech}. Arousal or activation (A) is the intensity of emotion provoked by the stimulus, ranges from sleepiness to excitement. The dominance (D) or power dimension refers to the degree of power or sense of control over the emotion \cite{gunes2010automatic}. This three-dimensional emotion model is known as VAD or PAD model 
\cite{mehrabian1974approach}. 

The concept of verbal communication is by conveying verbal words. However, some researchers reported that the use of non-verbal words, i.e., pause or silence, is needed for better human communication. Adding pause to emotional speech affects the recognition rate by human participants. Furthermore, silence and other disfluencies are not only useful for human communication but also can be effective cues for the computer to recognize human emotion \cite{tian2015recognizing}. 

An investigation on how speech pause length influences how listeners ascribe emotional states to the speaker has been done by authors in \cite{Tisljar-Szabo2014}. The author manipulated the length of speech pauses to create five variants of all passages. The participants were asked to rate the emotionality of these passages by indicating on a 1–6 point scale how angry, sad, disgusted, happy, surprised, scared, positive, and heated the speaker could have been. The data reveal that the length of silent pauses influences listeners in attributing emotional category to the speaker. Their findings argue that pauses play a relevant role in ascribing emotions and that this phenomenon might be partly independent of language.

Different from human to human communication, human to machine communication (or human-machine interaction, HMI) is a form of communication where humans interact with a variety of devices like sensors and actuators, or generally the computer. Although the silence aforementioned is useful for human emotion perception, it is still unclear whether it is useful or not for human to machine communication. One of the clue for this question is a study by Tian et al. \cite{tian2015recognizing}, \cite{moore2014word}, which used disfluencies and other non-verbal vocalizations as features for speech emotion recognition. Their results indicated that disfluencies and non-verbal vocalizations provide useful information overlooked by the other two types of features for emotion recognition: 
lexical and acoustic features. However, instead of using silences or pauses, they used filler pauses, fillers, stutters, laughter, breath, and sigh within an utterance to extract those features.

Instead of using silence feature, Atmaja and Akagi \cite{atmaja2019speech, atmaja2019b} removed silence within speech and extract acoustic features from the speech region after silence removal. Their results show an improvement of emotion category detection on an emotional speech dataset by utilizing silence removal and attention model. However, this method may slightly corrupts the speech fluency, because it generated a context of audio samples artificially.

The contribution of this paper is the investigation of the use of silence as a feature in automatic dimensional speech emotion recognition (SER). For each utterance, a number of frames are calculated and checked whether those frames can be categorized as silence. The fraction of the number of silence frames over total frames is measured as a silence feature. This silence feature is grouped with high statistical function (HSF), i.e., mean and standard deviation, of an acoustic feature set as the input to speech emotion recognition system. The comparison of HSF with and without silence feature can be used to determine the effect of silence feature on dimensional speech emotion recognition. The measure of comparison was given by the concordance correlation coefficient (CCC) \cite{lawrence1989concordance}. 

\section{Acoustic and silence features}
\subsection{Acoustic Feature Set}
Acoustic features are the input to an SER system. One of the acoustic feature sets proposed for SER is called Geneva Minimalistic Acoustic Parameter Set (GeMAPS), which is developed by Eyben et al. \cite{eyben2015geneva}. Those acoustic features extracted on frame-based processing are often called as Low-Level Descriptors (LLD). This frame-based processing is common in other speech processing applications. Other researchers \cite{schmitt2019continuous} proposed to ex tract functional features on certain lengths, e.g., 100 ms, 1 s, or per utterance/turn depend on the given labels. These functional features is often called as High-Level Statistical Functions (HSF). The reason for using HSF is to roughly describe the temporal and contour variations of different LLDs during certain period/utterance \cite{mirsamadi2017automatic}. Assuming that emotional content lies temporal variations rather than LLDs, HSFs may give a more accurate performance in determining emotional state from speech. Schmitt et al. suggested that using mean and standard deviation (std) from a set of acoustic features (GeMAPS) performed better than LLDs on speech emotion recognition \cite{schmitt2018deep}. We used these mean and std features, which are extracted per utterance from LLDs in GeMAPS feature set (2 $\times$ 23 features). To add those functionals, we proposed to use a silence feature, which is also extracted per utterance. The computation of a silence feature is explained below.

\begin{table}
\centering
\caption{GeMAPS feature \cite{eyben2015geneva} and its functionals used for 
         dimensional SER in this research.}
\begin{tabular}{l p{6cm}}
\hline
LLDs & loudness, alpha ratio, hammarberg index, spectral slope 0-500 Hz, spectral slope 
500-1500 Hz, spectral flux, 4 MFCCs, F0, jitter, shimmer, Harmonics-to-Noise Ratio (HNR), 
Harmonic difference H1-H2, Harmonic difference H1-A3, F1, F1 bandwidth, F1 
amplitude, F2, F2 amplitude, F3, and F3 amplitude. \\
\hline
HSFs & mean (of LLDs), standard deviation (of LLDs), silence \\
\hline
\end{tabular}
\label{tab:feature}
\end{table}

\subsection{Silence feature}
Silence, in this paper, is defined as the portion of the silence frames compared 
to the total frames in an utterance. In human communication, this portion of 
silence in speaking depends on the speaker's emotion. For example, a happy speaker 
may have fewer silences (or pauses) than a sad speaker. The portion of silence in 
an utterance can be calculated as

\begin{equation} \label{eq:silence}
    S = \frac{N_{s}}{N_{t}},
\end{equation}
where $N_s$ is the number of frames to be categorized as silence (silence frames), 
and $N_t$ is the number of total frames within an utterance. To be categorized as 
silence, a frame is checked whether it is less than a threshold, which is a 
multiplication of a factor with a root mean 
square (RMS) energy ($X_{rms}$). Mathematically, it can be formulated

\begin{equation}
    th = \alpha \times \overline{X_{rms}}
\end{equation}
and $X_{rms}$ is defined as

\begin{equation}
    X_{rms} = \sqrt{\frac{1}{n}\sum_{i=1}^{n}x[i]^2}
\end{equation}

\begin{figure}[htb]
\centering
\includegraphics[width=\linewidth]{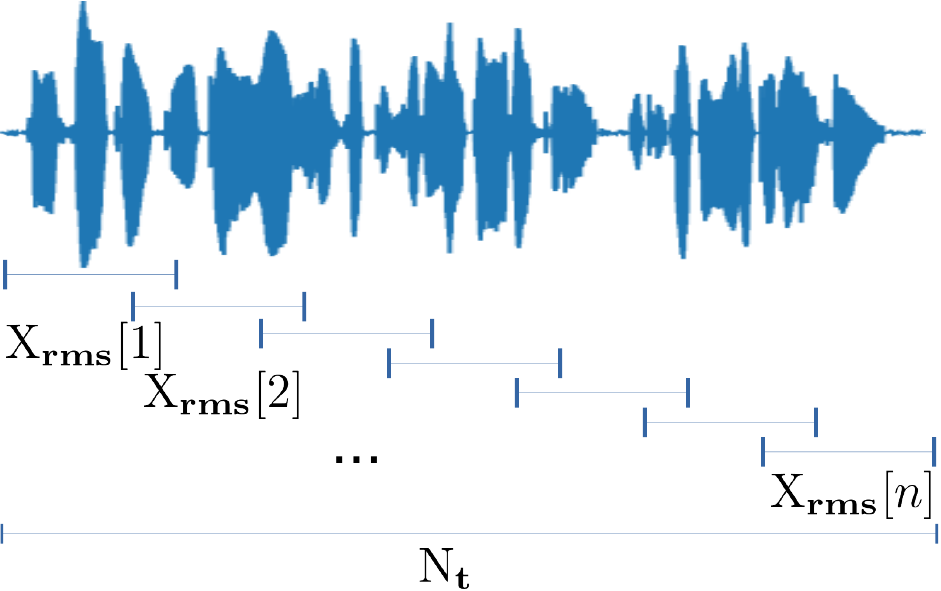}
\caption{The moving frame to calculate a silence feature}
\label{fig:silence_fig}
\end{figure}

These equations are similar to what is proposed in \cite{sahu2019multimodal}. The 
author of that paper used a fixed threshold, while we evaluated some factors of 
$\alpha$ to find the best factor for silence feature in speech emotion 
recognition. The equation \ref{eq:silence} to calculate the silence feature is also similar 
to the calculation of the disfluency feature proposed in \cite{moore2014word}. In 
that paper, the author divides the total duration of disfluency over the total 
utterance length on $n$ words. Fig. \ref{fig:silence_fig} shows the calculation of our 
silence feature. If $X_{rms}$ from a frame is below the $th$, then it is categorized 
as silence and follow the calculation of the equation \ref{eq:silence}.


\section{Experiments}

\subsection{Dataset}
The ``Interactive Emotional dyadic MOtion
CAPture" (IEMOCAP) database, collected by the
Speech Analysis and Interpretation Laboratory
(SAIL) at the University of Southern California (USC) was used to investigate 
the effect of silence feature on dimensional SER. 
This dataset consists of multimodal measurement of speech and gesture, 
including markers of the face, head, and hands, which provide detailed 
information about facial expressions and hand movements during a dyadic 
conversation. Among those modalities, only speech utterance is used. 
The total utterances are 10039 turns with three emotion attributes: arousal, 
valence, and dominance.  The average turn duration is 4.5 s with 
average 11.4 words per turn. The annotations are rated by at least two evaluators 
per utterance. The evaluators were USC students. We used emotion dimensions scores 
averaged from those two annotators as gold-standard labels in the experiments. 
The detail of that pilot study for developing the dataset can be found in \cite{busso2008iemocap}.

\subsection{Speech emotion recognition system}
SER is an attempt to make the computer recognize emotional states in speech. A deep neural 
network (DNN)-based SER is the common approach in recent days. Among numerous DNN methods, 
convolutional neural network (CNN) and LSTM are the most common \cite{schmitt2019continuous}, 
\cite{mirsamadi2017automatic}. 
We choose an LSTM-based dimensional SER due to its simplicity and the hardware support (CuDNN 
\cite{chetlur2014cudnn}). This architecture is a modification from the previous 
LSTM-based SER system reported in \cite{atmaja2019rnn} by enlarging the size of networks 
and using different parameters for multitask learning.

\begin{figure*}[h]
  \centering
  \includegraphics[width=\linewidth]{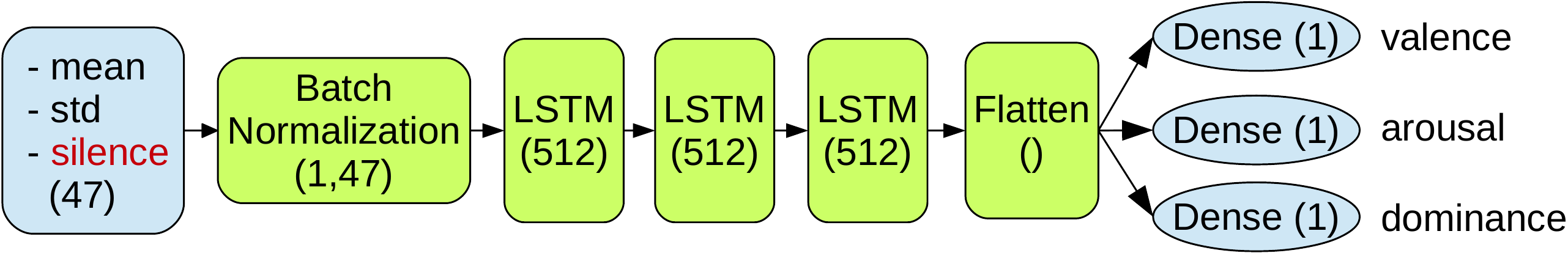}
  \caption{Structure of dimensional SER system to investigate the effect of silence features;
           the number inside the bracket represents the number of nodes/units.}
  \label{fig:ser_system}
\end{figure*}

For the input features, three sets of acoustic features are evaluated. These 
features are GeMAPS feature set (baseline); mean and std of GeMAPS (mean+std); and mean, std, and silence 
(mean+std+silence) features. The features in GeMAPS are extracted in 25 ms and 10 ms of the window and hop 
lengths using openSMILE feature extraction toolkit \cite{eyben2013recent}. 
Mean, std, and silence are extracted per utterance. The silence 
feature is extracted per utterance from 2048 samples of time frame length (128 ms) 
and 512 samples of hop length (32 ms) with 16000 Hz of sampling frequency. The implementation 
of silence feature computation was performed using LibROSA python package 
\cite{brian2019}. Those features are evaluated to the same architecture, shown in Fig. 
\ref{fig:ser_system}, which is implemented using Keras toolkit \cite{chollet2015keras}. 
Each frame shown in that figure represents a time frame to calculate $X_{rms}$ 
and to check whether it is a silence (if it is greater than $th$) or not.

The first layer on the dimensional SER system on that figure is the batch normalization layer.
This layer is intended to accelerate deep network training, as suggested in \cite{ioffe2015batch}. 
The size of the batch normalization layer depends on the input features. GeMAPS has 
the size of the nodes of ($3409 \times 23$) for IEMOCAP dataset, mean+std has a size of ($1 \times 46$), and mean+std+silence has a size 
of ($1 \times 47$). After a batch normalization layer, we stacked three LSTM layers (unidirectional, 512 nodes each) 
and flattened the output of the last LSTM layer. Three dense layers with each size of 1 are 
connected to Flatten layer to predict the degree of valence, arousal, and dominance. 
The degree of those emotion dimensions is a floating-point value ranges from [-1, 1], 
converted from the original 5-point scale.
The total size of the networks (trainable parameters) depends on the input features, about 10 million for GeMAPS input, and about 5 million for mean+std and mean+std+silence inputs.

For each input feature set, a number of 100 epochs were performed with earlystopping 
callbacks with a number of 10 patiences. This means, if the training process did not 
find an improvement of performance after 10 epochs, it will stop and save that 
best model for evaluation. To obtain a consistent/same result on each run, the 
same fixed random number is initiated at the top of the SER computer programs.

To measure the performance, a correlation measure, namely CCC, is used. This CCC 
is a measure of relation between prediction and true dimensional 
emotion degree (valence, arousal, dominance), which 
penalizes the score if the prediction shifts the true value. Instead of using  
a single value, we measure CCC for each emotion dimension. This method enables us 
to analyze which emotion dimension relates to specific features. The cumulative 
performance for all three dimensions can be given in an average of three 
CCC scores. The fair comparison can be performed between mean+std and mean+std+silence 
feature inputs, as it only has a difference in input size by a single value 
(46 vs. 47).

\section{Results and Discussion}
\subsection{Effect of silence feature on dimensional SER}
Although it is stated previously that the fair comparison could be made by comparing 
results from mean+std vs. mean+std+silence, for the sake of research continuity, 
the result from the previous reported result \cite{atmaja2019rnn} and GeMAPS feature are 
presented as baselines. Both kinds of research used the same SER architecture and the same input with 
different size of network (64 vs. 512 nodes for each LSTM layer). By using larger networks 
and different multitasking coefficients, 
an improvement of arousal has been obtained on GeMAPS feature input, 
while the CCC scores of both valences and dominance are similar. Our approach adopted multitask 
learning to train simultaneously valence, arousal, and dominance 
from \cite{atmaja2020multitask}. Here, the coefficients (weighting factors) used for valence, 
arousal, and dominance are 0.1, 0.5, and 0.4, respectively.  Table \ref{tab:result_ser} shows 
the obtained CCC score for each emotion dimension 
and its average score from different methods. 

Using HSFs of LLDs from GeMAPS, i.e., mean and std of 23 acoustic features, an 
improvement of valence was obtained. However, the CCC score of arousal and dominance  
decreased, although the average CCC score remains the same. This type of input feature 
(mean+std) has a smaller number of dimensions (1 $\times$ 46) compared to GeMAPS 
feature (3409 $\times$ 23). The size of the network of input with mean+std also 
about half of the network of GeMAPS input.

On the last method in Table \ref{tab:result_ser}, a silence feature was combined 
with std+mean resulting (1 $\times$ 47) of input size. This small modification 
leads to improvements in valence and arousal among other methods. A CCC score for this 
mean+std+silence input for dominance has decreased compared to GeMAPS, but 
slightly higher than mean+std. Both CCC scores on 
valence and arousal improved with 6\% and 17\% relative improvement. This 
result suggests that the silence feature affects arousal (activeness of speech) more 
than other dimensions. This finding may follow that humans tend to use 
more pauses in speech when they are sad and fewer pauses when they are happy. 

To extend this investigation, evaluation of the silence threshold factor ($\alpha$) 
was performed and discussed below.

\begin{table}[htb]
\centering
\caption{Results of dimensional emotion recognition by various methods measured in CCC score; 
         V: valence; A: arousal; D: dominance.}
\begin{tabular}{l c c c c}
\hline
Method & V  &   A   &   D   & Mean \\
\hline
Ref. \cite{atmaja2019rnn} &  0.11 & 0.43  & 0.36  & 0.30 \\
GeMAPS  &   0.118   &   0.536   &   \textbf{0.466}   &   0.373 \\
mean+std    & 0.201 &   0.476   &   0.435   &   0.371 \\
meant+std+silence   &   \textbf{0.214}   &   \textbf{0.561}   &   0.448   & \textbf{0.408}\\
\hline
\end{tabular}
\label{tab:result_ser}
\end{table}

\subsection{Evaluation of silence threshold factors} 
Most studies on silent pauses used threshold as one of the objects of study 
\cite{campione2002large}, \cite{rose2017silent}. 
Those studies categorized thresholds in silent pause into two groups: low threshold 
(200 ms) and high threshold (2000 ms). However, the definition of the threshold used 
here is different from those researches. The threshold in this research 
is defined as the upper-bound of RMS energy of a frame to be categorized 
as silence (equation (2)).

The silence threshold factor ($\alpha$) in equation (2) plays an important role in 
determining whether a frame belongs to the silence category. To investigate the 
effect of this factor on dimensional SER performance, we variate the $\alpha$ 
to 0.4, 0.3, 0.2, and 0.1. The result obtained in the previous Table \ref{tab:result_ser} 
with mean+std+silence input was obtained using $\alpha=0.3$.

\begin{figure}
\centering
\includegraphics[width=\linewidth]{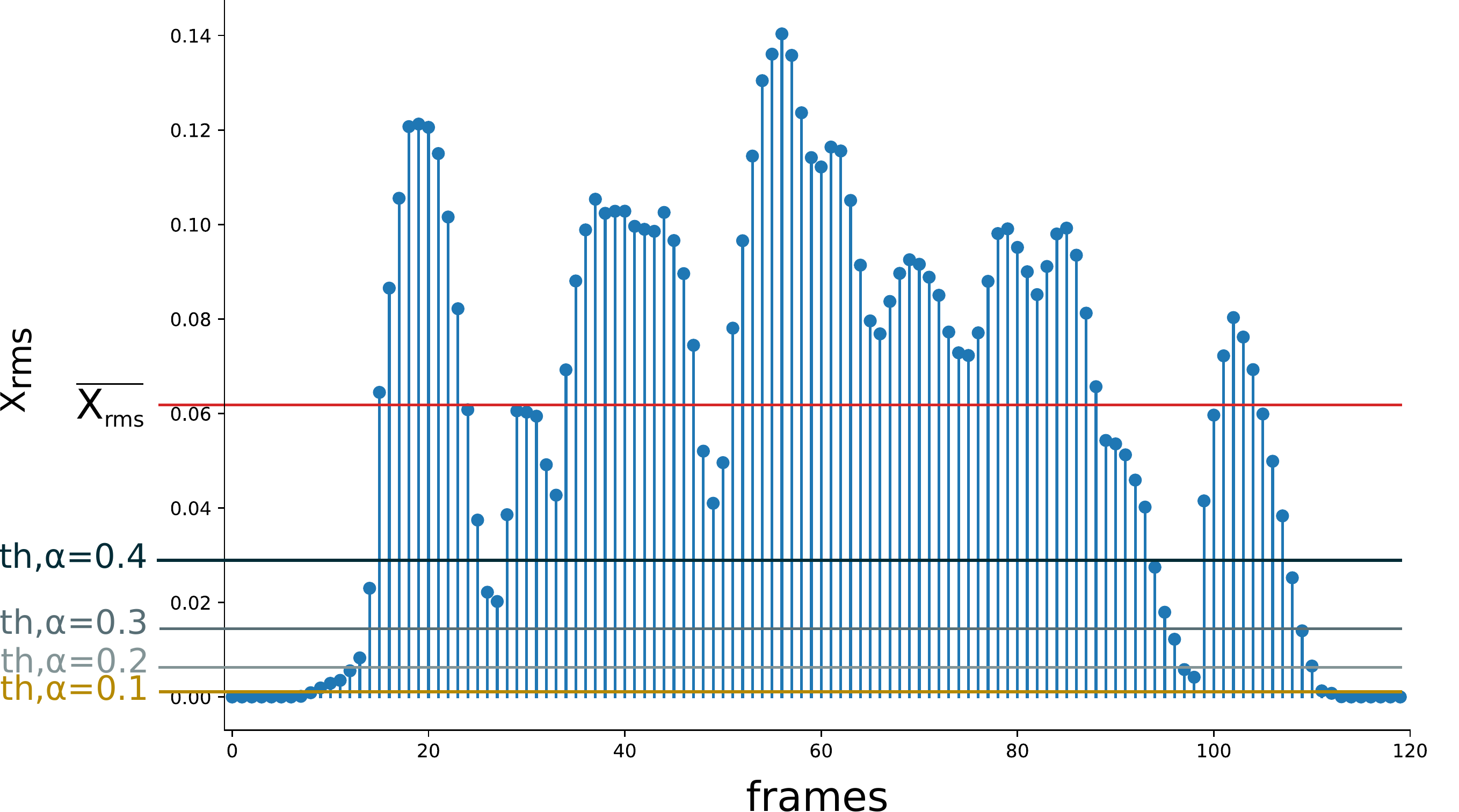}
\caption{RMS energy of corresponding frames with $\overline{X_{rms}}$ and 
threshold lines for different silence threshold factors.}
\label{fig:alpha}
\end{figure}

Fig. \ref{fig:alpha} shows the example of an utterance, its $X_{rms}$ of 
corresponding frame, $\overline{X_{rms}}$, and three lines of threshold using 
different silence threshold factors. As shown in that figure, using $\alpha=0.4$ 
may result in an incorrect decision to include speech as silence. However, using a low 
silence threshold factor, e.g., $\alpha=0.1$, leads to a smaller number of silence 
frames due to a tight filter. An evaluation to choose the proper factor is needed 
to obtain the optimal silence feature for dimensional SER.

\begin{figure}[htpb]
  \centering
  \includegraphics[width=\linewidth]{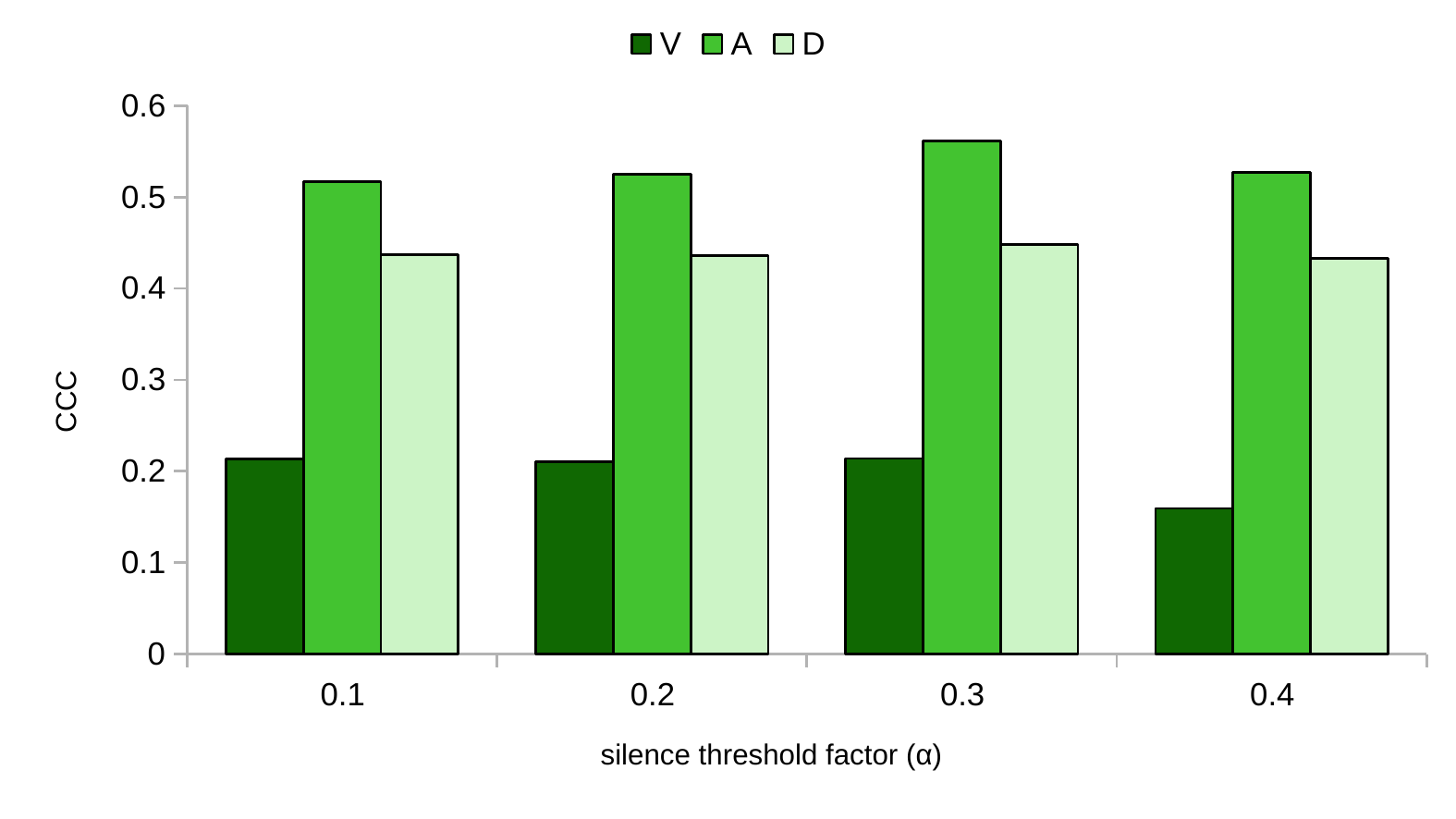}
  \caption{Evaluation of different silence threshold factor ($\alpha$) and its 
           impact on CCC score of valence, arousal, and dominance}
  \label{fig:alpha_result}
\end{figure}

Fig. \ref{fig:alpha_result} shows the effect of changing the silence threshold 
factor to the CCC score of valence, arousal, and dominance. Using a higher factor 
will impact on increasing the number of silence frames. On the other side, using 
a smaller factor will decrease the possibility to count a frame as a silence. As 
can be seen in that figure, the best CCC score was obtained using $\alpha=0.3$. 

The result shown on Fig. \ref{fig:alpha_result} also supports the finding that 
the silence affects the performance of predicting arousal. Using $\alpha=0.1$, 
$\alpha=0.2$, and $\alpha=0.3$ shows no difference on valence and dominance 
(0.21 and 0.43), but on arousal dimension. The CCC scores on arousal 
dimension are 0.51, 0.52, and 0.56 for $\alpha=0.1$, 
$\alpha=0.2$, and $\alpha=0.3$, respectively. Using $\alpha=0.4$ decreases 
the CCC scores of three emotional dimensions. 
This high silence threshold factor may select non-silence frames 
as silence frames. 
The average CCC scores for this $\alpha$ variation are 
0.389, 0.392, 0.408, and 0.373 for $\alpha=0.1$, 
$\alpha=0.2$, $\alpha=0.3$, and $\alpha=0.4$, respectively. 
These results also suggest that using improper silence threshold 
factor will decrease the performance of dimensional SER, especially on the 
arousal dimension.

\section{Conclusions}
In this paper, we investigate the effect of using silence feature on the 
dimensional speech emotion recognition. The result reveals that using mean+std+silence features affects 
the CCC score of predicted emotion degree compared to mean+std features. Using 
a proper factor of silence threshold, a remarkable improvement of CCC scores was obtained, 
particularly on arousal (activation) dimension. This can 
be explained that passiveness or activeness in speech, which reflected by 
number of pauses/silences in speech, contribute to arousal degree, as expected. 
On the other side, the use of improper silence 
threshold may decrease the performance of arousal. Using a fixed random number to 
initiate the computation of dimensional speech emotion recognition (same number for both 
mean+std and mean+std+silence for all architectures), the consistent results were 
obtained to support that finding on effect of silence on dimensional speech 
emotion recognition.

There are some issues which need to be confirmed for the future research. Although we 
obtained improvements in all emotion dimensions 
by using mean+std+silence from 
mean+std, the relationship between silence features with valence and dominance 
dimensions needs to be verified. The relation between positive and negative emotion 
dimensions with silence features is also meriting further study., e.g., more silences features with more 
valence, arousal, and dominance.

\bibliographystyle{IEEEtran}
\bibliography{mybib}

\begin{thebibliography}{10}
\providecommand{\url}[1]{#1}
\csname url@samestyle\endcsname
\providecommand{\newblock}{\relax}
\providecommand{\bibinfo}[2]{#2}
\providecommand{\BIBentrySTDinterwordspacing}{\spaceskip=0pt\relax}
\providecommand{\BIBentryALTinterwordstretchfactor}{4}
\providecommand{\BIBentryALTinterwordspacing}{\spaceskip=\fontdimen2\font plus
\BIBentryALTinterwordstretchfactor\fontdimen3\font minus
  \fontdimen4\font\relax}
\providecommand{\BIBforeignlanguage}[2]{{%
\expandafter\ifx\csname l@#1\endcsname\relax
\typeout{** WARNING: IEEEtran.bst: No hyphenation pattern has been}%
\typeout{** loaded for the language `#1'. Using the pattern for}%
\typeout{** the default language instead.}%
\else
\language=\csname l@#1\endcsname
\fi
#2}}
\providecommand{\BIBdecl}{\relax}
\BIBdecl

\bibitem{ibm_developer_2018}
J.~Sukis and L.~Lawrence, ``The human-to-machine communication model,''
  https://developer.ibm.com/articles/cc-design-cognitive-models-machine-learning/,
  Jan 2018, {A}ccessed: 15\--May\--2019.

\bibitem{scherer2000psychological}
K.~R. Scherer \emph{et~al.}, ``Psychological models of emotion,'' \emph{The
  neuropsychology of emotion}, vol. 137, no.~3, pp. 137--162, 2000.

\bibitem{ringeval2019avec}
F.~Ringeval, B.~Schuller, M.~Valstar, N.~Cummins, R.~Cowie, L.~Tavabi,
  M.~Schmitt, S.~Alisamir, S.~Amiriparian, E.-M. Messner \emph{et~al.}, ``Avec
  2019 workshop and challenge: state-of-mind, detecting depression with ai, and
  cross-cultural affect recognition,'' in \emph{Proceedings of the 9th
  International on Audio/Visual Emotion Challenge and Workshop}, 2019, pp.
  3--12.

\bibitem{moore2014word}
J.~D. Moore, L.~Tian, and C.~Lai, ``Word-level emotion recognition using
  high-level features,'' in \emph{International Conference on Intelligent Text
  Processing and Computational Linguistics}.\hskip 1em plus 0.5em minus
  0.4em\relax Springer, 2014, pp. 17--31.

\bibitem{jurafsky2018speech}
D.~Jurafsky and J.~H. Martin, \emph{Speech and language processing},
  3rd~ed.\hskip 1em plus 0.5em minus 0.4em\relax (Draft of September 11, 2018).
  Retrieved March, 2018.

\bibitem{gunes2010automatic}
H.~Gunes and M.~Pantic, ``Automatic, dimensional and continuous emotion
  recognition,'' \emph{International Journal of Synthetic Emotions (IJSE)},
  vol.~1, no.~1, pp. 68--99, 2010.

\bibitem{mehrabian1974approach}
A.~Mehrabian and J.~A. Russell, \emph{An approach to environmental
  psychology.}\hskip 1em plus 0.5em minus 0.4em\relax the MIT Press, 1974.

\bibitem{tian2015recognizing}
L.~Tian, C.~Lai, and J.~Moore, ``Recognizing emotions in dialogues with
  disfluencies and non-verbal vocalisations,'' in \emph{Proceedings of the 4th
  Interdisciplinary Workshop on Laughter and Other Non-verbal Vocalisations in
  Speech}, vol.~14, 2015, p.~15.

\bibitem{Tisljar-Szabo2014}
E.~Tislj{\'{a}}r-Szab{\'{o}} and C.~Pl{\'{e}}h, ``{Ascribing emotions depending
  on pause length in native and foreign language speech},'' \emph{Speech
  Communication}, vol.~56, no.~1, pp. 35--48, 2014.

\bibitem{atmaja2019speech}
B.~T. Atmaja and M.~Akagi, ``Speech emotion recognition based on speech segment
  using lstm with attention model,'' in \emph{2019 IEEE International
  Conference on Signals and Systems (ICSigSys)}.\hskip 1em plus 0.5em minus
  0.4em\relax IEEE, 2019, pp. 40--44.

\bibitem{atmaja2019b}
B.~T. {Atmaja}, K.~{Shirai}, and M.~{Akagi}, ``Speech emotion recognition using
  speech feature and word embedding,'' in \emph{2019 Asia-Pacific Signal and
  Information Processing Association Annual Summit and Conference (APSIPA
  ASC)}, Nov 2019, pp. 519--523.

\bibitem{lawrence1989concordance}
L.~Lawrence I-kuei, ``A concordance correlation coefficient to evaluate
  reproducibility,'' \emph{Biometrics}, pp. 255--268, 1989.

\bibitem{eyben2015geneva}
F.~Eyben, K.~R. Scherer, B.~W. Schuller, J.~Sundberg, E.~Andr{\'e}, C.~Busso,
  L.~Y. Devillers, J.~Epps, P.~Laukka, S.~S. Narayanan \emph{et~al.}, ``The
  geneva minimalistic acoustic parameter set (gemaps) for voice research and
  affective computing,'' \emph{IEEE Transactions on Affective Computing},
  vol.~7, no.~2, pp. 190--202, 2015.

\bibitem{schmitt2019continuous}
M.~Schmitt, N.~Cummins, and B.~Schuller, ``Continuous emotion recognition in
  speech--do we need recurrence?'' \emph{Training}, vol.~34, no.~93, p.~12,
  2019.

\bibitem{mirsamadi2017automatic}
S.~Mirsamadi, E.~Barsoum, and C.~Zhang, ``Automatic speech emotion recognition
  using recurrent neural networks with local attention,'' in \emph{2017 IEEE
  International Conference on Acoustics, Speech and Signal Processing
  (ICASSP)}.\hskip 1em plus 0.5em minus 0.4em\relax IEEE, 2017, pp. 2227--2231.

\bibitem{schmitt2018deep}
M.~Schmitt and B.~Schuller, ``Deep recurrent neural networks for emotion
  recognition in speech,'' in \emph{Proceedings DAGA}, vol.~44, 2018, pp.
  1537--1540.

\bibitem{sahu2019multimodal}
G.~Sahu, ``Multimodal speech emotion recognition and ambiguity resolution,''
  \emph{arXiv preprint arXiv:1904.06022}, 2019.

\bibitem{busso2008iemocap}
C.~Busso, M.~Bulut, C.-C. Lee, A.~Kazemzadeh, E.~Mower, S.~Kim, J.~N. Chang,
  S.~Lee, and S.~S. Narayanan, ``Iemocap: Interactive emotional dyadic motion
  capture database,'' \emph{Language resources and evaluation}, vol.~42, no.~4,
  p. 335, 2008.

\bibitem{chetlur2014cudnn}
S.~Chetlur, C.~Woolley, P.~Vandermersch, J.~Cohen, J.~Tran, B.~Catanzaro, and
  E.~Shelhamer, ``cudnn: Efficient primitives for deep learning,'' \emph{arXiv
  preprint arXiv:1410.0759}, 2014.

\bibitem{atmaja2019rnn}
B.~T. Atmaja, R.~Elbarougy, and M.~Akagi, ``{RNN}-based dimensional speech
  emotion recognition,'' in \emph{ASJ Autum Meeting}.\hskip 1em plus 0.5em
  minus 0.4em\relax Acoustical Society of Japan, 2019, pp. 743--744.

\bibitem{eyben2013recent}
F.~Eyben, F.~Weninger, F.~Gross, and B.~Schuller, ``Recent developments in
  opensmile, the munich open-source multimedia feature extractor,'' in
  \emph{Proceedings of the 21st ACM international conference on
  Multimedia}.\hskip 1em plus 0.5em minus 0.4em\relax ACM, 2013, pp. 835--838.

\bibitem{brian2019}
B.~McFee, V.~Lostanlen, M.~McVicar, A.~Metsai, S.~Balke, C.~Thomé, C.~Raffel,
  D.~Lee, F.~Zalkow, K.~Lee, and et~al., ``librosa/librosa: 0.7.1,'' Oct 2019.

\bibitem{chollet2015keras}
F.~Chollet \emph{et~al.}, ``Keras,'' \url{https://keras.io}, 2015.

\bibitem{ioffe2015batch}
S.~Ioffe and C.~Szegedy, ``Batch normalization: Accelerating deep network
  training by reducing internal covariate shift,'' in \emph{32nd International
  Conference on Machine Learning}, 2015, pp. 448--456.

\bibitem{atmaja2020multitask}
B.~T. Atmaja and M.~Akagi, ``Multitask learning and multistage fusion for
  dimensional audiovisual emotion recognition,'' in \emph{ICASSP, IEEE
  International Conference on Acoustics, Speech and Signal Processing
  Proceedings (to appear)}, 2020.

\bibitem{campione2002large}
E.~Campione and J.~V{\'e}ronis, ``A large-scale multilingual study of silent
  pause duration,'' in \emph{Speech prosody 2002, international conference},
  2002.

\bibitem{rose2017silent}
R.~Rose, ``Silent and filled pauses and speech planning in first and second
  language production,'' \emph{TMH-QPSR}, p.~49, 2017.

\end{thebibliography}
\end{document}